\documentclass[11pt]{article} 
\usepackage{hyperref} 
\pdfoutput=1 
 
\usepackage{hyperref} 
\pdfoutput=1 
 
\begin{document} 
 
\title{Atoms in the Surf:\\ 
  Molecular Dynamics Simulation of the Kelvin-Helmholtz Instability using
  9 Billion Atoms} 
 
\author{David F. Richards, Liam D. Krauss, \\
  William H. Cabot, Kyle J. Caspersen, Andrew W. Cook,\\
  James N.  Glosli, Robert E. Rudd, Frederick H. Streitz \\
  \\\vspace{6pt} Lawrence Livermore National Laboratory, \\
  Livermore, CA 94551, USA}
 
\maketitle 
 
%% The abstract (in this file, and that submitted as text to arXiv) should 
%% include the exact phrase 
%% "fluid dynamics video" or "fluid dynamics videos" 
 
\begin{abstract} 
  We present a fluid dynamics video showing the results of a 9-billion
  atom molecular dynamics simulation of complex fluid flow in molten
  copper and aluminum.  Starting with an atomically flat interface, a
  shear is imposed along the copper-aluminum interface and random atomic
  fluctuations seed the formation of vortices.  These vortices grow due
  to the Kelvin-Helmholtz instability.  The resulting vortical
  structures are beautifully intricate, decorated with secondary
  instabilities and complex mixing phenomena.

  This work performed under the auspices of the U.S. Department of
  Energy by Lawrence Livermore National Laboratory under Contract
  DE-AC52-07NA27344.
\end{abstract} 
 
% main text 
 
\section{Introduction} 

\begin{picture}(0,0)
  \unitlength\textwidth
  \put(1,-0.3){\makebox(0,0)[r]{LLNL-ABS-407874}}
\end{picture}%
In nature, fluids are comprised of atoms and molecules.  The complex
structures that form in fluid flow are the result of the collective
motions of those elementary constituents.  In simulation, computers
struggle to deal with the multitude of atoms needed to represent complex
fluid flows, but as computers become faster and more capable it is now
possible to perform molecular dynamics simulations sufficiently large to
reproduce fluid dynamics phenomena.

Our video, (%
\href{http://hdl.handle.net/1813/11528/KH9B_GFM_10_16_08A.m2v}{large} and %
\href{http://hdl.handle.net/1813/11528/KH9B_GFM_10_16_08A.m1v}{small}) %
shows the formation and evolution of vortices formed by shear flow in a
9-billion atom molecular dynamics simulation.  The simulation domain is
a quasi-2D geometry $12 \mu\textrm{m} \times 6 \mu\textrm{m} \times 2
\textrm{nm}$ with an equal number of aluminum and copper atoms at a
temperature of 2000~K, well above the melting temperature.  Energies and
forces between atoms are calculated using the EAM potential\cite{EAM} of
Mason, Rudd, and Sutton\cite{MSR}.  The simulation was performed on the
BG/L computer at LLNL using 212,000 CPUs.  The calculation required over
$36\times 10^6$~CPU hours or roughly 4 CPU millennia.  Further details of
the computational aspects of this work can be found in ref~\cite{gb}.

The simulation begins with laminar flow in which the copper at the
bottom of the image flows uniformly to the left and the aluminum at the
top flows to the right with an atomically flat interface between the
two.  The relative velocity of the two layers is 2000~m/s.  Color in the
video indicates the fraction of copper atoms (red is 100\% Cu; purple is
100\% Al).

The simulation runs for $10^6$ time steps or a total time of 2~ns during
which vortices develop and grow at the interface due to the
Kelvin-Helmholtz instability.  As time progresses we see the vortices
combine and grow as well as the formation of secondary instabilities.
Work to compare the results of this simulation to traditional
hydrodynamic methods is ongoing.

\end{document}